\documentstyle[aps,twocolumn,epsfig]{revtex}
\begin{document}
\title{Threshold  model  and  the  latest  NA50  data on $J/\psi$
suppression in Pb+Pb collisions}

\author{\bf A. K. Chaudhuri\cite{byline}}
\address{ Variable Energy Cyclotron Centre\\
1/AF,Bidhan Nagar, Kolkata - 700 064\\}
\maketitle
\begin{abstract}

Using the QGP motivated threshold model, where all the $J/\psi$'s
are  suppressed  above  a threshold density, we have analyzed the
latest version of the NA50 data on the centrality  dependence  of
the  $J/\psi$ over Drell-Yan ratio. The data are not well explain
in the model, unless the threshold density  is  largely  smeared.
Large  smeared threshold density effectively excludes creation of
any deconfined medium in the collision. \end{abstract} 

\section{Introduction}

In  relativistic  heavy  ion  collisions $J/\psi$ suppression has
been recognized as an important tool  to  identify  the  possible
phase transition to quark-gluon plasma. Because of the large mass
of  the  charm  quarks,  $c\bar{c}$ pairs are produced on a short
time scale. Their tight binding also makes them immune  to  final
state interactions. Their evolution probes the state of matter in
the  early  stage  of the collisions. Matsui and Satz \cite{ma86}
predicted that in presence of quark-gluon plasma  (QGP),  binding
of  $c\bar{c}$  pairs  into  a  $J/\psi$  meson will be hindered,
leading to the  so  called  $J/\psi$  suppression  in  heavy  ion
collisions  \cite{ma86}.  Over  the  years  several  groups  have
measured the $J/\psi$ yield in heavy ion collisions (for a review
of the data and the interpretations see Refs.  \cite{vo99,ge99}).
In  brief,  experimental  data  do show suppression. However this
could be attributed to the conventional nuclear absorption,  also
present in $pA$ collisions.

In  1998 NA50 collaboration \cite{na50b} published the results of
centrality dependence of $J/\psi$ suppression in 158 GeV/c  Pb+Pb
collisions.  Data  gave  the  first  indication  of the anomalous
mechanism  of  charmonium  suppression,  which  goes  beyond  the
conventional  suppression  in a nuclear environment. The ratio of
J/$\psi$ yield to that of Drell-Yan pairs decreases  faster  with
$E_T$  in  the  most  central collisions than in the less central
ones. It has been suggested that the  resulting  pattern  can  be
understood  in  a  deconfinement  scenario in terms of successive
melting of charmonium bound states \cite{na50b}. Later, the  data
were  analyzed  in  a  variety  of  models,  with  or without the
assumption      of       deconfining       phase       transition
\cite{bl00,ca00,ch01,ch02,qiu98}.

Recently, in Quark Matter 2002, NA50 collaboration presented {\em
preliminary}  analysis  of  the  $E_T$ dependence of the $J/\psi$
over  Drell-Yan  ratio,  obtained   in   the   2000   Pb+Pb   run
\cite{na50a}.  The suppression pattern is changed, presumably due
to changed  method  of  analysis.  In  1998,  NA50  collaboration
performed  two types of analysis, the standard analysis where the
$J/\psi$ and Drell-Yan cross sections were measured  directly  to
obtain  the ratio. In the other analysis, called the minimum bias
(MB) analysis, the Drell-Yan cross sections were replaced by  the
minimum bias cross section, to obtain,

\begin{equation}  \left [\frac{\sigma(J/\psi)}{\sigma(DY)} \right
]_{MB}= \left  [\frac{\sigma(J/\psi)}{\sigma(MB)}  \right  ]_{EX}
\left [\frac{\sigma(MB)}{\sigma(DY)} \right ]_{TH} \end{equation}

The  data  beyond  100  GeV  were  obtained  entirely from the MB
analysis \cite{na50b}. The latest NA50 data are obtained from the
standard analyzed only. Compared to 1998 data \cite{na50b},  2000
data  \cite{na50a}  are  flatter,  suppression  being more at low
$E_T$ and less at high $E_T$. However, the  suppression  obtained
is  still  anomalous  in the sense that normal nuclear absorption
model  fails  to  explain  it.  In  Quark   Matter   2002,   NA50
collaboration  also  presented  their  analysis  of  the  nuclear
absorption of $J/\psi$ in high statistics 450 GeV  pA  collisions
\cite{na50-a}.  They  estimated  the  $J/\psi$ nucleon absorption
cross section ($\sigma^{J/\psi N}_{abs}$)  in  the  framework  of
Glauber   model.   High   statistics   450   GeV  pA  data  yield
$\sigma^{J/\psi N}_{abs} = 4.4 \pm 1.0$  mb  \cite{na50-a}.  They
also  estimate  a common $\sigma^{J/\psi N}_{abs}$ from latest pA
and  NA38  200  GeV/c  S+U  data   \cite{na38},   $\sigma^{J/\psi
N}_{abs}$
=4.4 $\pm$ 0.5 mb.
The  extracted  absorption cross section is much smaller than the
earlier value of 6.4 $\pm$ 0.8 mb extracted from fit  to  earlier
NA50  data  \cite{na50-b} or 7.1 $\pm$ 3.0 mb obtained from a fit
to NA38  S+U  data  \cite{na38}.  Within  error,  the  S+U  cross
sections are compatible with pA cross sections.

The   changed   suppression  pattern  in  the  latest  NA50  data
\cite{na50a},  along   with   the   high   statistics   pA   data
\cite{na50-a},  with  the implication of smaller $J/\psi$-nucleon
absorption cross  section,  necessitated  re-examination  of  the
models,  which  were successful in explaining the earlier version
of the NA50 data. Recently we have analyzed the latest NA50  data
in  the  QCD  based  nuclear  absorption  model \cite{ch03a}. The
parameters  of  the  model  were  fixed  from  the  recent   high
statistics  pA  data.  Without any free parameter the model could
explain the latest NA50 data.  Capella  et  al  \cite{ca03}  also
analyzed   the   data  in  the  comover  model.  $J/\psi$-nucleon
absorption cross section was fixed at 4.5  mb.  The  latest  NA50
data   are   explained  with  comover-$J/\psi$  absorption  cross
section, $\sigma_{co}$=0.65 mb.

Blaizot  et  al  \cite{bl00} proposed the QGP motivated threshold
model  to  explain  the  earlier  version  of   the   NA50   data
\cite{na50b}.  To mimic the onset of deconfining phase transition
above  a  critical  energy  density  and  subsequent  melting  of
$J/\psi$'s, $J/\psi$ suppression was linked with the local energy
density.  If  the  energy  density at the point where $J/\psi$ is
formed, exceeds a critical  value  ($\varepsilon_c$),  $J/\psi$'s
disappear.  The critical energy density was then related with the
(transverse) density and its value was obtained from a fit to the
NA50 experimental data \cite{na50b}.  The  NA50  data  were  well
fitted with critical (threshold) density $n_c$=3.7-3.75 $fm^{-2}$
and  $\sigma^{J/\psi  N}_{abs}$=6.4  mb.  The  model  needs to be
tested against the latest NA50 data  \cite{na50a},  with  changed
suppression pattern, taking into consideration the new data on pA
collisions \cite{na50-a}.

Aim  of  the  present  paper  is  to analyze the latest NA50 data
\cite{na50a}  on  the  centrality  dependence  of  $J/\psi$  over
Drell-Yan  ratio, in the threshold model \cite{bl00}. Plan of the
paper is as follows: In section 2, we present a brief description
of the threshold model. In section 3, NA50 data are  analyzed  in
the model. Summary and conclusions are drawn in section 4.

\section{$J/\psi$ suppression in threshold model}

The details of the threshold model could be found in \cite{bl00}.
In  addition  to  the  'conventional' nuclear absorption, Blaizot
{\em et al.}  \cite{bl00}  introduced  an  anomalous  suppression
factor,  $S_{anom}$,  such  that  all  the $J/\psi$'s are totally
suppressed  above  a  critical  (threshold)  density  $n_c$.  The
$J/\psi$  cross  section  at  an  impact parameter ${\bf b}$ as a
function of $E_T$ is then written as,

\begin{eqnarray} \label{blaizot}
\frac{d^3\sigma^{J/\psi}}{dE_Td^2b} = &&\sigma^{J/\psi}_{NN}
\int d^2s T^{eff}_A({\bf s}) T^{eff}_B({\bf b-s})\\
&& S_{anom}({\bf b,s}) P(b,E_T) \nonumber
\end{eqnarray}

\noindent  where $T^{eff}(b)$ is the effective nuclear thickness,
$T^{eff}({\bf  b})=\int_{-\infty}^{\infty}  dz  \rho({\bf   b},z)
exp(-\sigma^{J/\psi  N}_{abs}  \int_z^{\infty} dz\prime \rho({\bf
b},z\prime)$. For the density $\rho$, we use a Woods-Saxon form

\begin{equation}
\rho(r)=\frac{\rho_0}{1+exp((r-R)/a)}, \hspace{1cm}\int d^3r \rho(r)=A
\end{equation}

In  \cite{bl00},  Blaizot  et  al used $R=1.1A^{1/3}=6.52 fm$ and
$a$=.53  fm.  However,  nuclear  absorption   has   a   sensitive
(exponential)  dependence on density and it is better to use more
realistic values for the radius and the diffusiveness parameters.
For $Pb$ we use,  $R=6.624  fm$  and  $a$=0.549  fm  \cite{de74}.
$\sigma^{J/\psi  N}_{abs}$  is  the  $J/\psi$-nucleon  absorption
cross section for which a value of =4.4 mb \cite{na50-a} is used.

In   Eq.\ref{blaizot},  $P(b,E_T)$  is  the  $E_T-b$  correlation
function for which a Gaussian form is used \cite{bl00},

\begin{equation}\label{5}
P(b,E_T) \propto exp(-(E_T-qN_p(b))^2/2q^2aN_p(b)),
\end{equation}

\noindent where $N_p(b)$ is the number of participant nucleons at
impact  parameter  b.  $a$  and  $q$  are  parameters  related to
dispersion and average transverse energy.  For  Pb+Pb  collisions
the  parameters  are,  $a$=1.27  and  $q$=0.274  GeV respectively
\cite{bl00}.

$S_{anom}({\bf   b,s})$  in  Eq.\ref{blaizot}  is  the  anomalous
suppression factor introduced by Blaizot et al \cite{bl00}.  They
considered  two  forms  for  $S_{anom}$.  Assuming  that  all the
$J/\psi$'s get suppressed above a threshold density ($n_c$),  the
anomalous suppression factor can be written as,

\begin{equation}  \label{s1}  S_{anom}({\bf b,s}) =\Theta (n({\bf
b,s})-n_c) \end{equation}

\noindent   where   $n$   is  the  transverse  density  (Eq.3  of
\cite{bl00}). In ref.\cite{bl00} it was seen that  if  the  theta
function  is smeared by a small amount, at the expense of another
parameter, such that suppression is gradual rather  than  abrupt,
the  quality  of fit to data improves considerably. This was done
by writing,

\begin{equation}   \label{s2}   S_{anom}({\bf  b,s})  =  0.5  [1-
tanh(\lambda (n({\bf b,s})-n_c))] \end{equation}

In both the forms, effect of $E_T$ fluctuations at a fixed impact
parameter  is  taken into account by rescaling the density as, $n
\rightarrow n E_T/<E_T>({\bf  b})$  \cite{bl00}.  Threshold model
 parameters
$n_c$  and $\lambda$ can be obtained by fitting the latest NA50
data on centrality dependence of $J/\psi$ over Drell-Yan ratio.

The Drell-Yan pairs do not suffer any final state interaction and
the  cross section at impact parameter ${\bf b}$ as a function of
$E_T$ can be calculated as,

\begin{equation}         \frac{d^3\sigma^{DY}}{dE_Td^2b}        =
\sigma^{DY}_{NN} \int d^2s T_A({\bf s}) T_B({\bf b-s})
 P(b,E_T) \end{equation}

\begin{figure}[h]
\centerline{\psfig{figure=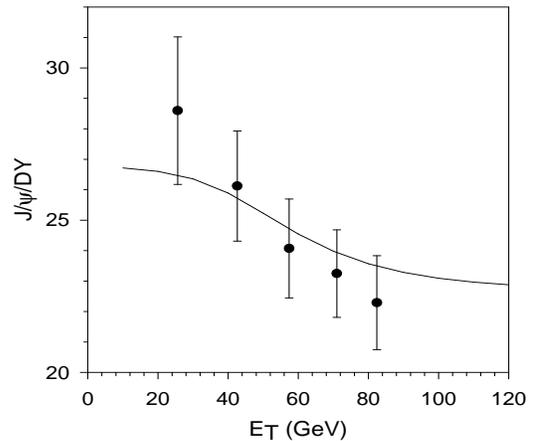,height=10cm,width=8cm}}
 \vspace{-2.5cm}   \caption{The   NA38  data  on  the  centrality
dependence of $J/\psi$ over Drell-Yan ratio in 200 GeV/c S+U
collisions. The
line  is  the  Glauber   model   of   nuclear   absorption   with
$\sigma^{J/\psi N}_{abs}$=4.4 mb. For the $E_T-B$ correlation, we
have  used  Gaussian  form  Eq.\ref{5},  with  a=3.2  and q=0.74 GeV
\protect \cite{vo99}. } \end{figure}

\section{Comparison with latest NA50 data}

For  comparison  with  NA50  experimental  data on the centrality
dependence of $J/\psi$ over Drell-Yan ratio,  normalizing  factor
$N=B_{\mu\mu}\sigma^{J/\psi}_{NN}/\sigma^{DY}_{NN}$ is needed. It
can  also be considered as a fitting parameter and obtained along
with  the  threshold  model  parameters,  $n_c$  and   $\lambda$.
However,  normalizing  factor  and threshold model parameters are
correlated. Higher  normalizing  factor  can  be  compensated  by
increasing  the threshold density. Noting that the ratio of Pb+Pb
to   S+U   normalizations   is   equal   to   $1.051\pm    0.026$
\cite{na50-a,ca03},  NA38  data  on  the centrality dependence of
$J/\psi$ over Drell-Yan ratio in 200 GeV S+U data \cite{na38} can
be fitted to obtain the normalizing factor for S+U collisions and
can be rescaled to obtain the same for Pb+Pb collisions.  In  S+U
collision, deconfinement transition is not expected. The observed
suppression   is   due   to   nuclear   absorption  only.  Fixing
$\sigma^{J/\psi N}_{abs}$=4.4 mb, NA38 data  are  fitted  in  the
conventional  nuclear  absorption model to obtain the normalizing
factor. In Fig.1, NA38 data along with the best fit obtained with
$B_{\mu\mu}\sigma^{J/\psi}_{NN}/\sigma^{DY}_{NN}$=39.65 is shown.
Within the error bars, the data are well explained in the nuclear
absorption model. Normalizing factor for Pb+Pb collision is  then
obtained  by scaling the normalizing factor for S+U collisions by
1.051 \cite{na50-a}.

We  may mention that recently, Capella et al \cite{ca03} analyzed
the  S+U  data  in  the  comover  model.   With   $\sigma^{J/\psi
N}_{abs}$=4.5  mb,  data  allow  for a small comover interaction,
$\sigma_{co}$=0.65.                    For                    the
$B_{\mu\mu}\sigma^{J/\psi}_{NN}/\sigma^{DY}_{NN}$   they  used  a
value of 47, obtained from a fit to latest NA50 Pb+Pb data in the
comover model. However, as we find the data are well explained in
terms of nuclear absorption only, use of comover  interaction  in
S+U collision is debatable.

\begin{figure}[h]
\centerline{\psfig{figure=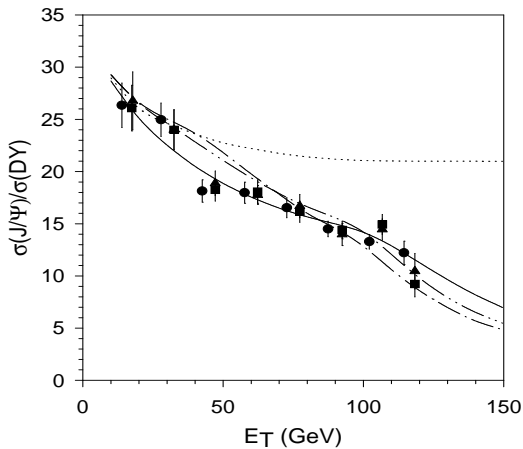,height=10cm,width=8cm}}
\vspace{-2.5cm}  \caption{The  latest NA50 data on the centrality
dependence of $J/\psi$ over Drell-Yan ratio in Pb+Pb  collisions.
The  dotted  line is the Glauber model of nuclear absorption with
$\sigma^{J/\psi N}_{abs}$=4.4 mb. The  dash-dotted  line  is  the
ratio  in  the  threshold  model  with  $n_c$=3.61 $fm^{-2}$. The
dash-dot-dot line  is  obtained  with  $n_c$=3.72  $fm^{-2}$  and
$\lambda$=2  $fm^2$  (fixed).  The  solid line is obtained with ,
$n_c$=3.82 $fm^{-2}$ and $\lambda$=0.77 $fm^{2}$.} \end{figure}

In  Fig.2,  latest  NA50  data  \cite{na50a}  on  the  centrality
dependence of $J/\psi$ over Drell-Yan ratio in Pb+Pb  collisions,
is  shown.  Just to show that the latest data are also anomalous,
we have shown the Glauber model calculation with  $\sigma^{J/\psi
N}_{abs}$=4.4  mb  (the  dashed  line).  Only for very peripheral
collisions, the Glauber model  of  nuclear  absorption  fits  the
data.   For  more  central  collisions,  it  produces  much  less
suppression than the data exhibit. In Fig.1, the dash-dotted line
is the best fit obtained to the  data  in  the  threshold  model,
without  any  smearing  of  the  threshold  density.  We obtained
threshold density $n_c$=3.61 $\pm$ 0.06$fm^{-2}$.  The  value  is
less  than  value  $n_c$=3.7 $fm^{-2}$, obtained by Blaizot et al
\cite{bl00} from the earlier version of the data, mainly  due  to
lesser value of the $J/\psi$-Nucleon absorption cross section. To
produce  similar  suppression,  $\sigma^{J/\psi  N}_{abs}$  being
less, anomalous suppression has to  increase.  Interestingly,  we
find  that  the  threshold model with a single parameter does not
give a proper description to the data. In the intermediate  range
of $E_T$, agreement with data is not good. Smearing the threshold
density  by  a small amount do not improve the quality of fit. In
Fig.2, the  dash-dot-dot  line  is  obtained  with  smearing  the
threshold  density  by  a  small  amount  ($\lambda$  fixed  at 2
$fm^2$). The best fit to the data is then obtained with threshold
density,$n_c$=3.72  $fm^{-2}$.  Again  the  data  are  not   well
explained.  In  Fig.2, the solid line is the best fit obtained to
the data varying both the parameters. We obtain, $n_c$=3.82 $\pm$
0.09  $fm^{-2}$ and $\lambda$=0.77 $\pm$ 0.14 $fm^{2}$. The model
then reproduces the data through out the $E_T$ range. Small value
of  $\lambda$  required  for  good  fit  to  data  indicate  that
considerable  smearing  of  the threshold density is required for
proper description of the NA50 data. The anomalous suppression is
not abrupt but increases gradually  with  density.  The  analysis
suggest  that  with  $\sigma^{J/\psi N}_{abs}$=4.4 mb, unless the
threshold density is largely smeared, threshold model do not give
a proper description of the latest NA50 data  on  the  centrality
dependence of the $J/\psi$ suppression.

\section{summary and conclusions}

To  summarize,  the latest NA50 data on the centrality dependence
of $J/\psi$ over Drell-Yan ratio in Pb+Pb collisions are analyzed
in the QGP motivated threshold model. In the threshold model,  in
addition  to  the  conventional  nuclear absorption, an anomalous
suppression is introduced, such that above a  threshold  density,
all  the $J/\psi$'s are absorbed. To be consistent with latest pA
data  on  $J/\psi$  absorption,  we  have  used   $\sigma^{J/\psi
N}_{abs}$=4.4  mb.  Threshold  model with a single parameter, the
threshold density ($n_c$) donot give a proper description of  the
centrality  dependence of $J/\psi$ over Drell-Yan ratio. The best
fit  to  data  is  obtained  with  threshold  density  $n_c$=3.61
$fm^{-2}$  fails  to  explain  the data in the intermediate $E_T$
range. If the threshold density is  smeared  at  the  expense  of
another  parameter ($\lambda$) the model could explain the latest
NA50 data with $n_c$=3.82 $fm^{-2}$ and $\lambda$=0.77  $fm^{2}$.
Small  value  of $\lambda$=0.77 $fm^2$ indicate that for a proper
description to the NA50 data, the threshold  density  has  to  be
smeared  considerably. Thus onset of anomalous suppression is not
sudden,  rather  gradual.  Over  a  density  range   of   2.4-5.2
$fm^{-2}$,  anomalous  suppression factor change from 0.9 to 0.1.
Originally, threshold model was devised to mimic the  melting  of
$J/\psi$'s  in a deconfining medium. With nominal smearing of the
threshold density, the essence of the model is  not  lost.  Large
smearing  of the threshold density, as required to fit the latest
NA50 data, effectively excludes formation of deconfining  medium.
The  medium  where  the anomalous suppression is taking place, do
not melt the $J/\psi$ suddenly, but rather gradually,  more  like
in a nuclear/comover environment.

\end{document}